\documentclass[conference]{IEEEtran}
\IEEEoverridecommandlockouts
\usepackage{cite}
\usepackage{amsmath,amssymb,amsfonts}
\usepackage{algorithmic}
\usepackage{graphicx}
\usepackage{subfigure}
\usepackage{textcomp}
\usepackage{xcolor}
\usepackage{makecell}
\usepackage{pifont}
\def\BibTeX{{\rm B\kern-.05em{\sc i\kern-.025em b}\kern-.08em
    T\kern-.1667em\lower.7ex\hbox{E}\kern-.125emX}}
\begin{document}

\title{ATA-Cache: Contention Mitigation for GPU Shared L1 Cache with Aggregated Tag Array


}

\author{Xiangrong Xu, Liang Wang, Limin Xiao, Lei Liu, Xilong Xie, Meng Han, Hao Liu \\
State Key Laboratory of Software Development Environment \\
School of Computer Science and Engineering \\ 
Beihang University, Beijing 100191, China \\
\{xxr0930@buaa.edu.cn\} }

\maketitle

\begin{abstract}

GPU shared L1 cache is a promising architecture while still suffering from high resource contentions. We present a GPU shared L1 cache architecture with an aggregated tag array that minimizes the L1 cache contentions and takes full advantage of inter-core locality. The key idea is to decouple and aggregate the tag arrays of multiple L1 caches so that the cache requests can be compared with all tag arrays in parallel to probe the replicated data in other caches. The GPU caches are only accessed by other GPU cores when replicated data exists, filtering out unnecessary cache accesses that cause high resource contentions. The experimental results show that GPU IPC can be improved by 12\% on average for applications with a high inter-core locality.

\end{abstract}


\section{Introduction}
GPUs are widely used in various applications such as machine learning, high performance computing, etc., because of their massive parallelism and high energy efficiency\cite{choukse2020buddy,kwon2021tensor,sun2022gtuner,kim2018gpu,wang2022a2}. 
Cache plays a vital role in GPU architecture due to its ability to address the memory wall problems\cite{wulf1995hitting}. In a typical GPU two-level cache architecture, each GPU core has a private L1 cache, and all GPU cores share a banked L2 cache. 
However, the conventional GPU cache hierarchy results from the inefficiencies that the same cache lines are replicated in multiple L1 caches when requested by multiple cores (i.e., inter-core locality), causing a low utilization of the expensive on-chip memory\cite{li2015inter}. 

Prior studies attempt to share L1 cache for multiple cores in GPU to better utilize the inter-core locality\cite{dublish2016cooperative,ibrahim2019analyzing,ibrahim2020analyzing,ibrahim2021analyzing}. However, the shared L1 cache architectures still suffer from various side effects such as higher resource contentions, longer L1 access time, etc., which negatively degrade the overall GPU performance. 
Existing shared L1 cache architectures for GPU can be classified into \textbf{remote-sharing L1 cache}\cite{dublish2016cooperative,ibrahim2019analyzing} and \textbf{decoupled-sharing L1 cache}\cite{ibrahim2020analyzing,ibrahim2021analyzing} based on the address mapping policy of the L1 cache. 

For remote-sharing L1 cache, each L1 cache is still closely coupled to a GPU core and mapped to the entire address space. In case of an L1 cache miss, the core sends probe requests to remote L1 caches (L1 caches of other cores) and decides whether to access the L2 cache based on the responses from the remote caches. A fatal drawback is that the requests to access the L2 cache have to wait for a long time period until determining whether the remote caches have the replicated data, which increases the critical path for L2 cache access and hurts GPU performance. Moreover, these probe requests take up a lot of NoC resources, and the remote caches require additional resources to process these probe requests.

Decoupled-sharing L1 cache architectures\cite{ibrahim2020analyzing,ibrahim2021analyzing} are proposed to solve the problem of resource contentions caused by additional probe requests in the remote-sharing L1 cache. A couple of cores are clustered and the L1 caches are decoupled from the cores. Each L1 cache in the cluster is exclusively mapped to a slice of the address range, so that all L1 caches in the cluster are mapped to the entire address space. Requests from different GPU cores that access the same cache line are mapped to the same L1 cache. 
However, decoupled-sharing L1 cache faces serious cache bank conflicts when multiple GPU cores access the same cache line simultaneously. The serialization of access requests from different GPU cores becomes a key bottleneck affecting the overall GPU performance.

To address these problems, we propose ATA-Cache, a GPU shared L1 cache architecture with an aggregated tag array. ATA-Cache can mitigate the L1 cache contentions and take full advantage of inter-core locality. 
In the proposed design, the L1 cache tag array is decoupled from the cache data array. The tag arrays of GPU cores in a cluster are aggregated together. 
Access requests from GPU cores can be compared to all tag arrays in parallel to find the replicated data in other L1 caches without probing requests, thus leveraging inter-core locality with minimal NoC contentions. In order to reduce the bank conflicts of the L1 cache, we adopt remote-sharing cache data, i.e., each L1 cache is mapped to the entire address space. If there is no inter-core locality, each GPU core can still access its own L1 cache data in parallel. 
ATA-Cache can take full advantage of the inter-core locality to improve GPU performance and achieve a 12\% IPC improvement on high inter-core locality applications. For applications with poor inter-core locality, there is no performance impairment due to sharing.   




\section{Background and Related Work}

\subsection{GPU Cache Architecture}
A typical GPU architecture contains multiple cores, each of which contains a private, dynamically partitioned GPU L1 cache and shared memory. GPU core accesses the L2 cache via NoC in case of an L1 cache miss. The GPU L2 cache is memory-side and shared by multiple GPU cores, which means that its access latency is typically several times higher than that of the L1 cache\cite{zhao2019adaptive}.
Due to the high L2 cache latency, many research efforts have been devoted to improving GPU performance by increasing L1 cache hit rate\cite{tripathy2021paver,baruah2020valkyrie,ibrahim2020analyzing,ibrahim2021analyzing,li2019efficient}.

Due to the GPU's private L1 cache design, when multiple GPU cores request data from the same cache line, the replicated data is loaded into each private L1 cache of the cores. The replicated data causes a low utilization of the expensive on-chip memory. The inter-core locality has been explored by sharing the L1 cache for multiple cores. In other words, when a request misses in the L1 cache, it can access replicated data in other L1 caches instead of going to the L2 cache, thus improving the cache hit rate. Based on the address mapping policy of the L1 cache, existing shared L1 cache architectures for GPU can be classified into remote-sharing L1 cache\cite{dublish2016cooperative,ibrahim2019analyzing} and decoupled-sharing L1 cache\cite{ibrahim2020analyzing,ibrahim2021analyzing}. 

\begin{figure} 
	\centering 
	\subfigure[Remote-sharing L1 cache]{
		\label{Fig remote-sharing L1 cache_1}
		\includegraphics[width=0.48\linewidth]{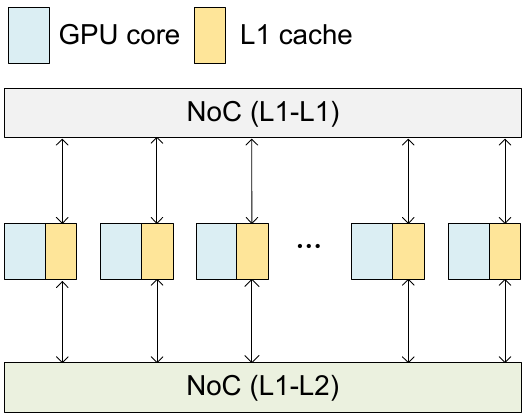}}
	\subfigure[Decoupled-sharing L1 cache]{
		\label{Fig decoupled-sharing L1 cache_1}
		\includegraphics[width=0.48\linewidth]{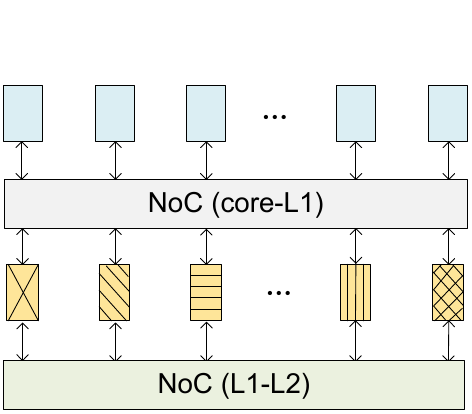}}
	\caption{Remote-sharing L1 cache and decoupled-sharing L1 cache design. Different textures of the L1 cache represent different address ranges.}
	\label{share L1 cache}
\end{figure}

\begin{figure}[t]
\centerline{\includegraphics[width=0.5 \textwidth]{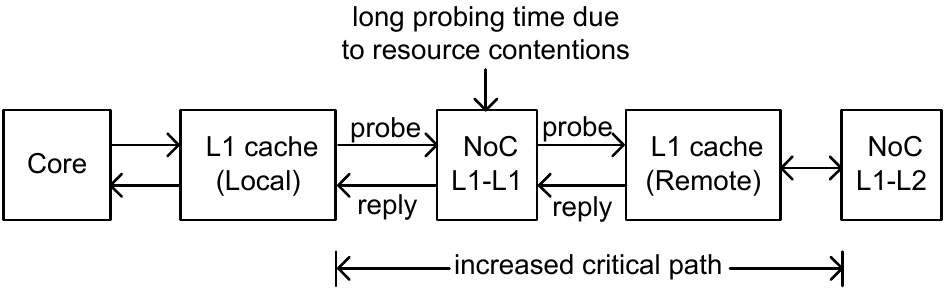}}
\caption{Request access process for remote-sharing L1 cache.}
\label{Fig remote-sharing L1 cache disadvantage}
\end{figure}

\subsection{Remote-sharing L1 Cache}
Figure \ref{Fig remote-sharing L1 cache_1} illustrates the remote-sharing L1 cache design. As shown in Figure \ref{Fig remote-sharing L1 cache_1}, each L1 cache is still closely coupled to a GPU core and mapped to the entire address space. Each L1 cache is not only connected to each partition of the L2 cache but also connected to other L1 caches through NoC (mesh, X-bar, etc.). For example, Dublish et al.\cite{dublish2016cooperative} propose L1 Cooperative Caching Network, a lightweight ring network that connects L1 caches in GPU cores. When a request misses in the L1 cache, the metadata of the request is first pushed into the ring network to be passed to other caches. The other caches receive the request, compare it with their own tag array, and if it hits, pass the data to the requesting GPU core using the cache network. Ibrahim et al.\cite{ibrahim2019analyzing} use a mesh network to connect the L1 caches and introduce a prediction mechanism when L1 caches probe remote data, thus reducing unnecessary probe requests, reducing the bandwidth pressure on the NoC, and improving the request response time.

To summarize, as shown in Figure \ref{Fig remote-sharing L1 cache disadvantage}, in the remote-sharing cache, missed requests need to access other L1 caches before going to the L2 cache, which increases the critical path for L2 cache access. What is worse, serious NoC resource contentions during accessing other L1 caches can significantly increase the L1 cache latency. These challenges make it difficult for remote-sharing cache to make better use of inter-core locality because of resource contentions overhead, and can even hurt GPU performance on applications with poor inter-core locality.


\begin{figure}
\centerline{\includegraphics[width=0.5 \textwidth]{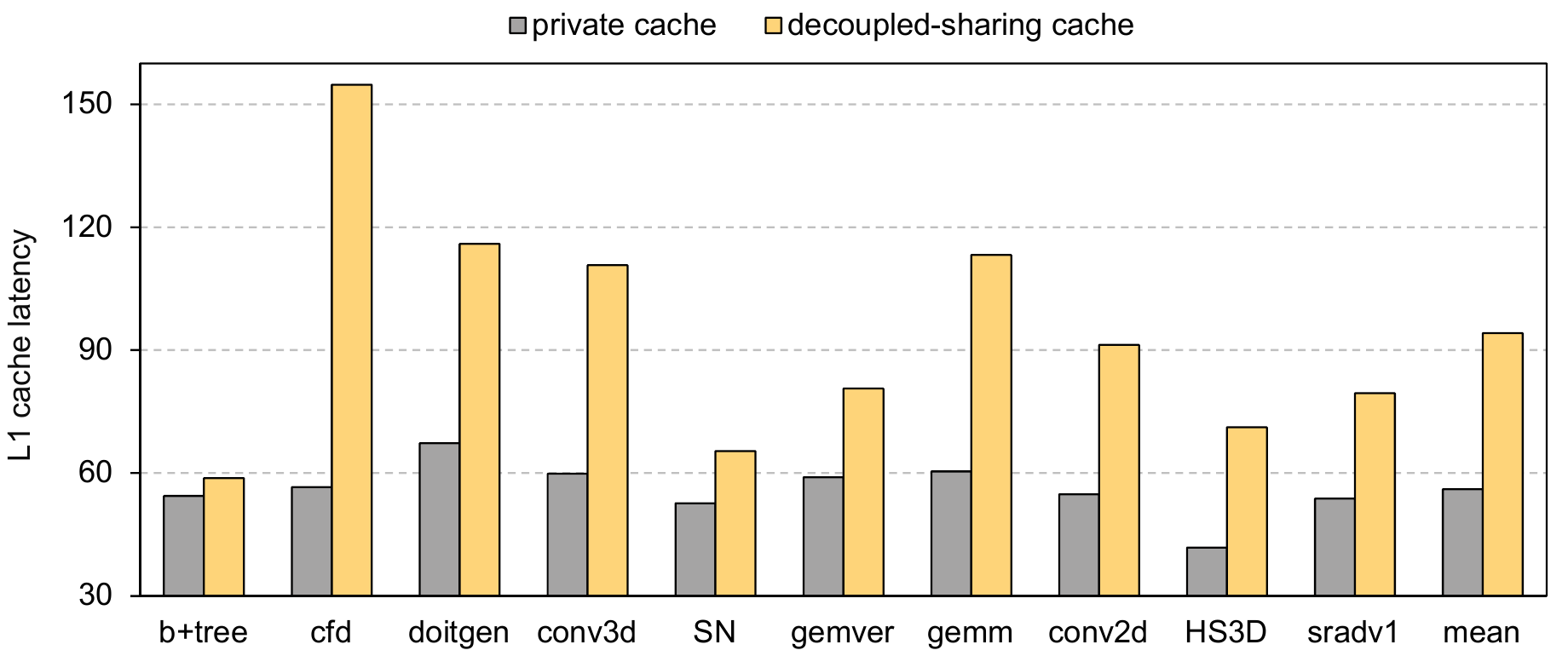}}
\caption{L1 cache latency for private cache and decoupled-sharing cache.}
\label{L1 cache latency for instructions private and decouple}
\end{figure}



\subsection{Decoupled-sharing L1 Cache}

As shown in Figure \ref{Fig decoupled-sharing L1 cache_1}, in the decoupled-sharing cache design, the L1 caches are decoupled from the GPU cores and they no longer have a one-to-one correspondence. A couple of cores are clustered with the L1 caches shared in the cluster. Each L1 cache is  exclusively mapped to a slice of the address range. 
Requests from different GPU cores access the L1 cache according to the address mapping rule. Decoupled-sharing L1 cache is first proposed in \cite{ibrahim2020analyzing}, where each L1 cache remains inside the GPU core but is mapped to a different address range. The recent work\cite{ibrahim2021analyzing} removes the L1 cache from inside the GPU core and connects it to the GPU core via the NoC, which is more conducive to sharing. The decoupled-sharing cache does not need to probe the replicated data of the remote cache (replicated data in all GPU cores are mapped to the same cache). However, requests from multiple GPU cores are often mapped to the same cache bank at the same time, which can lead to cache bank conflicts and serialization of parallel requests.


 \begin{table*}[]
\centering
\caption{Landspace of GPU shared L1 cache.}
\label{table: share vs private}
\resizebox{\textwidth}{!}{
\begin{tabular}{ccccccc}
\hline
\multicolumn{1}{|c|}{Cache design}            & \multicolumn{1}{c|}{L1 hit rate (\ding{115})} & \multicolumn{1}{c|}{\makecell{IPC for high \\ inter-core locality(\ding{115})}} & \multicolumn{1}{c|}{\makecell{IPC for poor \\ inter-core locality(\ding{115})}} & \multicolumn{1}{c|}{\makecell{L1 cache \\ latency(\ding{116})}} & \multicolumn{1}{c|}{\makecell{L2 cache bandwidth \\demand (\ding{116})}} & \multicolumn{1}{c|}{\makecell{Resource contentions \\ due to sharing(\ding{116})}} \\ \hline
\multicolumn{1}{|c|}{Private cache}           & \multicolumn{1}{c|}{\ding{73}}           & \multicolumn{1}{c|}{\ding{73}}                                & \multicolumn{1}{c|}{\ding{73}\ding{73}\ding{73}}                              & \multicolumn{1}{c|}{\ding{73}\ding{73}\ding{73}}              & \multicolumn{1}{c|}{\ding{73}}                         & \multicolumn{1}{c|}{N/A}                                 \\ \hline
\multicolumn{1}{|c|}{Remote-sharing cache}    & \multicolumn{1}{c|}{\ding{73}\ding{73}\ding{73}}         & \multicolumn{1}{c|}{\ding{73}\ding{73}}                               & \multicolumn{1}{c|}{\ding{73}}                                & \multicolumn{1}{c|}{\ding{73}}                & \multicolumn{1}{c|}{\ding{73}\ding{73}\ding{73}}                       & \multicolumn{1}{c|}{\ding{73}}                                   \\ \hline
\multicolumn{1}{|c|}{Decoupled-sharing cache} & \multicolumn{1}{c|}{\ding{73}\ding{73}\ding{73}}         & \multicolumn{1}{c|}{\ding{73}\ding{73}}                               & \multicolumn{1}{c|}{\ding{73}}                                & \multicolumn{1}{c|}{\ding{73}}                & \multicolumn{1}{c|}{\ding{73}\ding{73}\ding{73}}                       & \multicolumn{1}{c|}{\ding{73}}                                   \\ \hline
\multicolumn{1}{|c|}{ATA-Cache}               & \multicolumn{1}{c|}{\ding{73}\ding{73}\ding{73}}         & \multicolumn{1}{c|}{\ding{73}\ding{73}\ding{73}}                              & \multicolumn{1}{c|}{\ding{73}\ding{73}\ding{73}}                              & \multicolumn{1}{c|}{\ding{73}\ding{73}\ding{73}}              & \multicolumn{1}{c|}{\ding{73}\ding{73}\ding{73}}                       & \multicolumn{1}{c|}{\ding{73}\ding{73}\ding{73}}                                 \\ \hline
\multicolumn{7}{l}{\ding{115}: the higher the metric, the better;
\ding{116}: the lower the metric, the better.}                   
\end{tabular}}
\vspace{-1em}
\end{table*}
Figure \ref{L1 cache latency for instructions private and decouple} shows the L1 cache latency for private cache and decoupled-sharing cache. The decoupled-sharing cache, despite of higher hit rate, incurs much longer L1 latency than the private cache due to severe cache resource contentions. When multiple GPU cores access the same cache bank simultaneously, the resource contentions lead to request serialization, thus becoming a bottleneck for the overall performance. 
Therefore, to address the serious resource contention problem in the decoupled-sharing cache, it becomes imperative to study a shared L1 cache that better exploits inter-core locality.

\subsection{Comparison of Cache Architectures}

Table \ref{table: share vs private} presents a comparison between the existing GPU cache architectures and the proposed cache architecture. Since the remote-sharing cache, decoupled-sharing cache, and ATA-Cache can  take advantage of inter-core locality, they all have a higher L1 cache hit rate and lower L2 bandwidth demand than the private cache. However, both remote-sharing cache and decoupled-sharing cache face resource contentions due to sharing, making it difficult for them to exploit the benefits of sharing on applications with high inter-core locality. Moreover, in applications with poor inter-core locality, they are even inferior to private cache architecture due to high contentions. ATA-Cache outperforms existing shared cache architectures in both applications with high inter-core locality and applications with low inter-core locality due to lower resource contentions.

\section{Design and Implementation}
In this section, we present the design of ATA-Cache, which can leverage inter-core locality to enhance GPU performance while the resource contentions are mitigated.

\begin{figure}[t]
\centerline{\includegraphics[width=0.5 \textwidth]{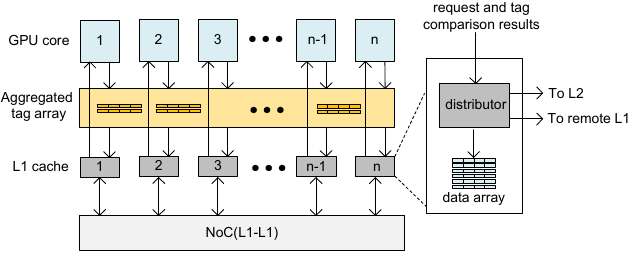}}
\caption{ATA-Cache design.}
\label{Fig:ATA-Overview}
\vspace{-1em}
\end{figure}

\subsection{Overview of ATA-Cache}\label{3A}
Figure \ref{Fig:ATA-Overview} provides a high-level overview of our shared cache design with an aggregated tag array. To allow multiple cores to share the L1 caches, the L1 caches are decoupled from the GPU cores. Each decoupled L1 cache is named the local cache of the GPU core, while the L1 caches of other GPU cores are named remote caches. Each GPU core accesses its local cache via a one-to-one connection after comparing tags and accesses remote caches from the local cache via the crossbar. For example, in Figure \ref{Fig:ATA-Overview}, GPU core 0 can access cache 0 (local cache) directly, or cache 1 (remote cache) through a crossbar. 

To utilize the data in the remote cache, we decouple the tag array in each cache and aggregate them together into an aggregated tag array. Requests from the core can be compared with multiple decoupled tag arrays in parallel to probe the replicated data in the remote cache. In contrast to remote-sharing cache, ATA-Cache completes replicated data probing when comparing tags, without sending probe requests to other caches and waiting for responses. 

Different from the decoupled-sharing cache, each L1 cache is mapped to the entire address space, avoiding additional contentions for the same cache bank. Each GPU core's request is first handled by a distributor in its local cache, which determines whether to access the data array of the local L1 cache or remote L1 caches, or even the L2 cache.  

For applications with low inter-core locality, each GPU core accesses its local cache data array in parallel. In this scenario, ATA-Cache is almost equivalent to the private cache, so that there are no additional cache bank conflicts. For applications with high inter-core locality, the remote cache is accessed only when it has replicated data, thus filtering out unnecessary cache accesses. Note that when multiple cores access the same remote cache, cache bank conflicts can still happen, which are much slighter because unnecessary cache accesses are filtered out.

When a request is sent from the GPU core, it first enters the aggregated tag array. The request gets the location of the data in the cache after comparing the request address tag with the cache tags in the aggregated tag array. The request then goes to the request distributor in the local L1 cache. In the request distributor, the request chooses to go to the data array of the local cache, a remote cache, or the L2 cache, depending on the results of the tag comparison. Finally, the response data of the request is sent back to the GPU core. 

We detail the design in the aggregated tag array and L1 cache in Section \ref{3B} and Section \ref{3C}.

\subsection{Aggregated Tag Array Design}\label{3B}
\begin{figure}[]
\centerline{\includegraphics[width=0.48 \textwidth]{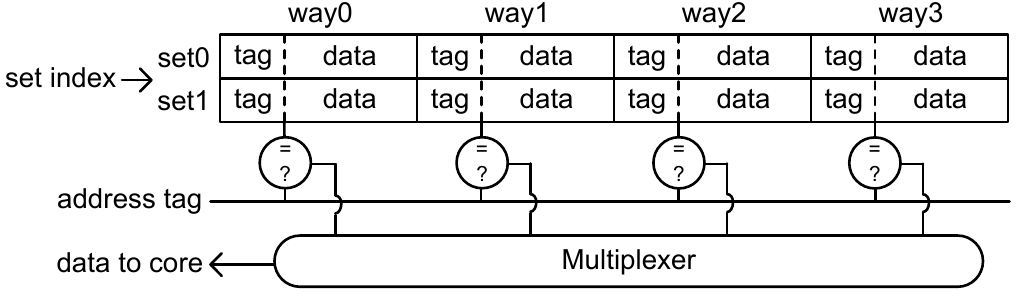}}
\caption{The tag array architecture for the conventional private L1 cache.}
\label{Fig: traditional tag array}
\end{figure}

\begin{figure*}[t]
\centerline{\includegraphics[width=0.85 \textwidth]{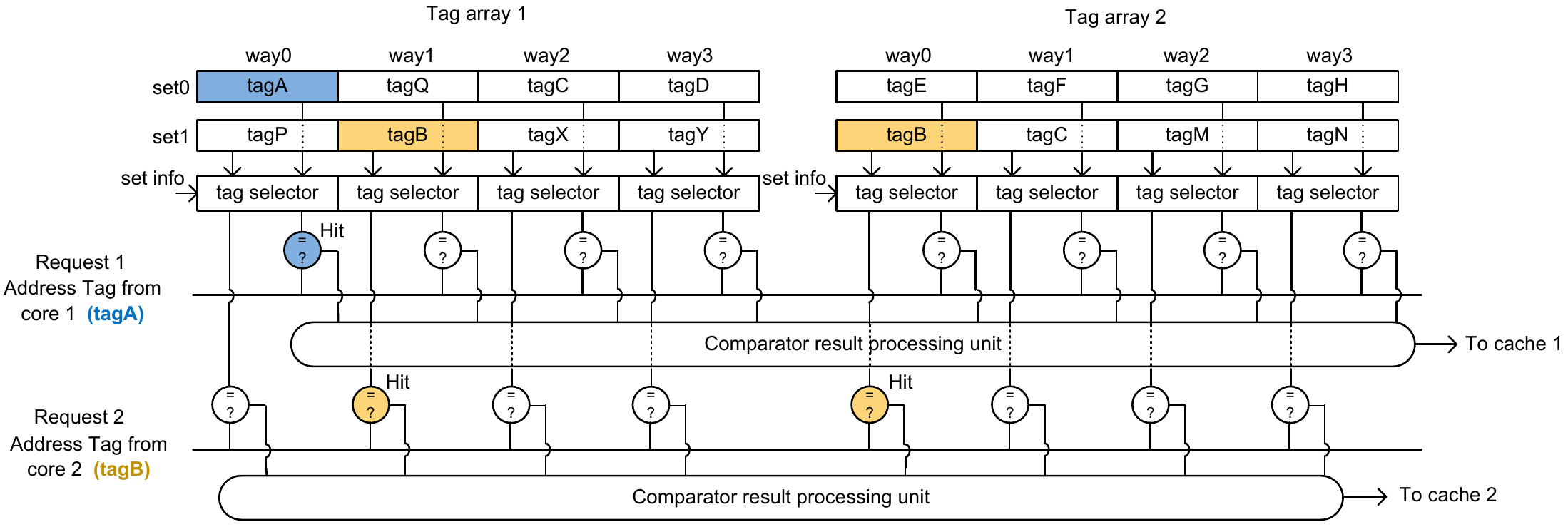}}
\caption{Aggregated tag array design. We show a parallel comparison of a request tag with the tags in the two tag arrays.}
\label{Fig:tag_array}
\vspace{-1em}
\end{figure*}
The tag array architecture for the conventional private L1 cache is shown in Figure \ref{Fig: traditional tag array}, in which the tag array and data array are closely coupled. 
When the request arrives at the tag array, the request address is first decoded to the cache set index and address tag. Then a set is selected according to the set index, and all the cache tags in it are input to the comparators and compared with the address tag in parallel. Finally, according to the comparison results, a multiplexer is used to select the required data and send it to the GPU core.

To implement the aggregated tag array, as shown in Figure \ref{Fig:tag_array}, we take two 4-way tag arrays as an example. In the design of aggregated tag array, we mainly need to address two problems. 
First, the requests face the bank conflict problem when they need to be compared with the tags in multiple sets in the same tag array. Therefore, in our design, each set in the tag array is located on a separate bank, so that there is no bank conflict even if different sets are accessed at the same time. Second, since different sets in the tag array are selected by requests from different cores, it faces the problem of how to send the tags to the correct comparators. Taking the two requests in Figure \ref{Fig:tag_array} as an example, Req-1 from Core-1 needs to be compared with the tags in set 0, and Req-2 from Core-2 needs to be compared with the tags in set 1. For the problem of how to send the tags in two sets to the corresponding comparators, we design the tag selector. The input to the tag selector is the cache tags in different sets, and the selection signal is the set index of each request. The selector sends the cache tag to the comparator corresponding to each request separately. After solving the above two problems, the aggregated tag array can compare requests from different cores with the tags in all tag arrays in parallel to get the location of the requested data in all L1 caches.

\textbf{Working Example.} We now introduce a working example of the aggregated tag array. In Figure \ref{Fig:tag_array}, Core-1 and Core-2 send two requests, called Req-1 and Req-2, respectively. The set index of Req-1 is 0 and the address tag is tagA, while the set index of Req-2 is 1 and the address tag is tagB. Then both sets in the tag arrays are activated and the tags are sent to the tag selectors. The tag selectors send the cache tags in set 0 to the comparators associated with Req-1 and send the cache tags in set 1 to the comparators associated with Req-2. After comparison, Req-1 hits in Tag array 1 and misses in Tag array 2, so the result of comparing Req-1 with the aggregated tag array is [1,0]. Similarly, the result of comparing Req-2 with the aggregated tag array is [1,1], because the address of Req-2 is hit in both Tag array 1 and Tag array 2.

\begin{figure}[t]
	\centering 
	\subfigure[Req hits in remote cache]{
		\label{Fig Remote cache hits and is available}
		\includegraphics[width=0.3\linewidth]{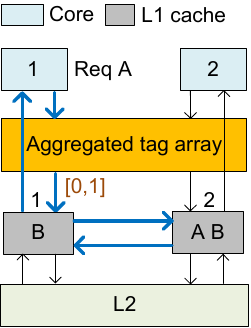}}
	\subfigure[Req hits in remote cache and local cache]{
		\label{Fig Local cache hits}
		\includegraphics[width=0.3\linewidth]{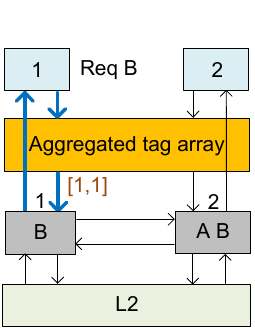}}
    \subfigure[Req misses in all L1 caches]{
		\label{Fig All L1 caches have missed}
		\includegraphics[width=0.3\linewidth]{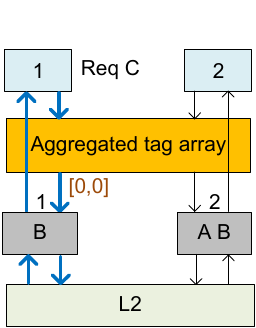}}
	\caption{Three cases of request distribution.}
	\label{Working Example}
\vspace{-1em}
\end{figure}
\subsection{L1 Cache Design}\label{3C}
The L1 cache without the tag array contains the request distributor, data array, etc. The request distributor receives the request tag comparison results and selects the target caches for the requests. As shown in Figure \ref{Working Example}, for the three different comparison results, requests go to different caches to access the data. 

First,  as shown in Figure \ref{Fig Remote cache hits and is available}, Core-1 sends a request for data A, which arrives at the aggregated tag array. After comparing the aggregated tag array, the output is [0,1], indicating that the request should be sent to cache 2. After fetching the data, the request returns to Cache 1, fills the data into cache 1, and returns to Core-1.
Second, as shown in Figure \ref{Fig Local cache hits},  since data B is available in both caches, the aggregated tag array is compared and the output value is [1,1]. In this case, we give priority to accessing the local cache of the GPU core. The request directly accesses the data array in cache 1 and then brings the data back to the GPU core. Finally, Figure \ref{Fig All L1 caches have missed} shows the case where all L1 caches are missed and the request needs to be sent to the L2 cache via NoC.

It is possible that the design of the decoupled tag array would result in data not being available immediately after comparing the tags. In the data array, a 1-bit dirty bit is used to indicate whether the data block has been changed. When a read request goes to the remote cache to fetch data, a write request in the remote cache modifies that data, and the dirty bit should be set to true. If the remote data is modified, the request needs to go to the L2 cache, although the probability of this happening is very low,  and has almost no impact on performance.

It's worth noting that to avoid introducing additional difficulties regarding cache coherency, for write requests we only process them in the local cache of the request's source core. By doing so, we do not need to change the GPU cache coherency mechanism. When handling non-data cache requests, such as share memory requests, texture requests, atomic operations, etc., our design is consistent with that of the private cache.

\section{Experimental Evaluation}
We have implemented our proposed design in GPGPU-sim v4.0\cite{khairy2020accel}. The detailed configurations of our modeled GPU are shown in Table \ref{table_config}. The GPU contains 30 SIMT cores, which are divided into three clusters. We faithfully simulate the competition and cost of the cache access process. We select ten applications from three benchmark suites (Rodinia 3.1\cite{che2009rodinia},  Tango\cite{karki2019tango}, Polybench\cite{Polybench}) and classify these applications into the high inter-core locality and low inter-core locality applications based on the amount of replicated data across all cores. 
\begin{table}[t]
    \renewcommand{\arraystretch}{1.3}
    \caption{Configuration parameters of the simulated GPU.}
    \label{table_config}
    \raggedleft
    \begin{tabular}{|l|l|}
        \hline
        Paremeters & Value \\
        \hline
        GPU core Features & 30 SIMT cores,1.365GHz, 4 GTO schedulers/core\\
        \hline
       L1 Caches/Core &\begin{tabular}[l]{@{}l@{}} 64KB 64-way sector L1 data cache,\\4banks, LRU, 128B cache line, \\32B sector size, latency = 32 cycles 
        \end{tabular}\\
        \hline
        L2 Cache & \begin{tabular}[l]{@{}l@{}} 16-way 128KB/memory sub partition\\(3MB in total),  128B cache line size, \\32B sector size, latency = 188 cycles 
        \end{tabular}\\ 
        \hline
        Memory Model&\begin{tabular}[l]{@{}l@{}} 12 Memory Controllers, 16 DRAM-banks,\\ 3.5GHz memory clock, $t_{CL}$ = 20, $t_{RP}$ = 20, \\ $t_{RC}$ = 62, $t_{RAS}$ = 50,  $t_{CCD}$ = 4, $t_{RCD}$ = 20,\\ $t_{RRD}$ =10, $t_{CDLR}$ = 9, $t_{WR}$ = 20 
        \end{tabular}\\
        \hline
        Interconnect & \begin{tabular}[l]{@{}l@{}}30 × 24 crossbar topology, 1.365GHz interconnect \\ clock, 40B flit size, in buffer limit 512, \\out buffer limit 512, iSLIP Arbiteration type
        \end{tabular}\\ 
        \hline
    \end{tabular}
\vspace{-1em}
\end{table}

\subsection{Overall Performance}\label{4A}
\begin{figure}[t]
\centerline{\includegraphics[width=0.5 \textwidth]{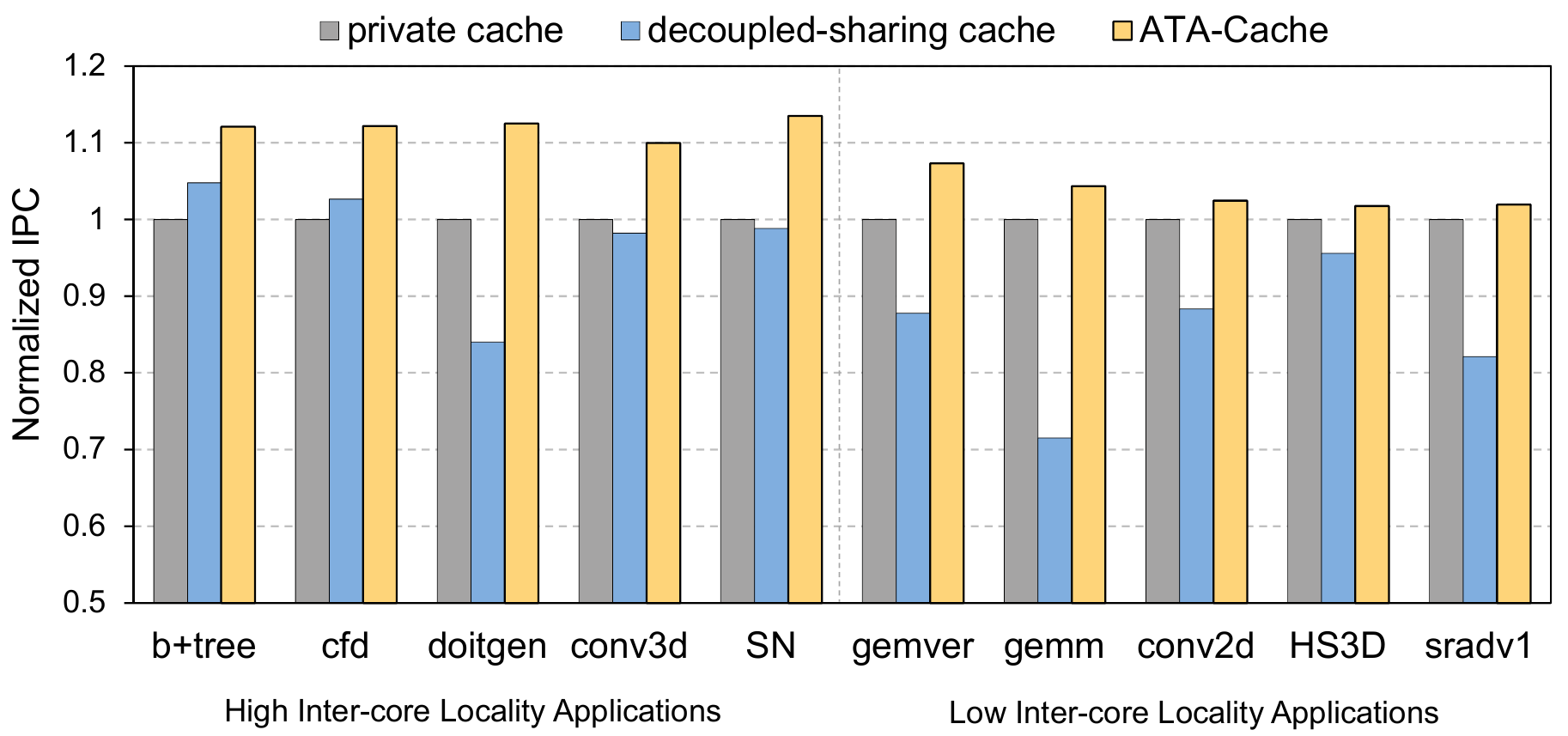}}
\caption{Illustrating the benefits of the ATA-Cache in terms of overall IPC (normalized to private cache).}
\label{Fig: Normalized IPC}
\vspace{-1em}
\end{figure}

We use IPC (instructions per cycle) to represent the performance. Figure \ref{Fig: Normalized IPC} shows the performance of different L1 cache architectures on high inter-core locality applications and low inter-core locality applications. The decoupled-sharing cache only performs better than the private L1 cache on two applications(b+tree and cfd) and it is even worse than the private L1 cache on other applications. As shown in Figure \ref{Fig: Normalized IPC}, for high inter-core locality applications, decoupled-sharing L1 cache can reduce the data replication of L1 cache in different GPU cores, but it leads to lower gains in exploiting inter-core locality because multiple requests are mapped to the same cache bank. For b+tree and cfd, the performance profit from the decoupled-sharing cache design is higher than the performance loss due to cache bank conflicts, resulting in GPU performance improvement. For doitgen, conv3d, and SN, the decoupled-sharing cache is less effective than the private cache due to severe resource contentions. The ATA-Cache has distinct performance improvements on all ten applications, especially on high inter-core locality applications. For five high inter-core locality applications, ATA-Cache reduces resource contentions caused by sharing while exploiting inter-core locality. In ATA-Cache requests can access data from the cache of other cores, which reduces response time and significantly improves performance with IPC increased by 12.0\% on average. 

For low inter-core locality applications, the performance loss of decoupled-sharing cache caused by resource contentions is higher than the performance improvement from exploiting inter-core locality, which is the reason why it performs more poorly than private cache.
At the same time, for low inter-core locality applications, requests from each core almost only can access the corresponding L1 cache in parallel, which reduces bank conflicts. Therefore, ATA-Cache outperforms decoupled-sharing cache by 22.9\% on average for applications with a low inter-core locality.

\subsection{Performance per Kernel}\label{4B}
\begin{figure}[t]
	\centering 
	\subfigure[SN]{
		\label{kernel performance SN}
		\includegraphics[width=0.48\linewidth]{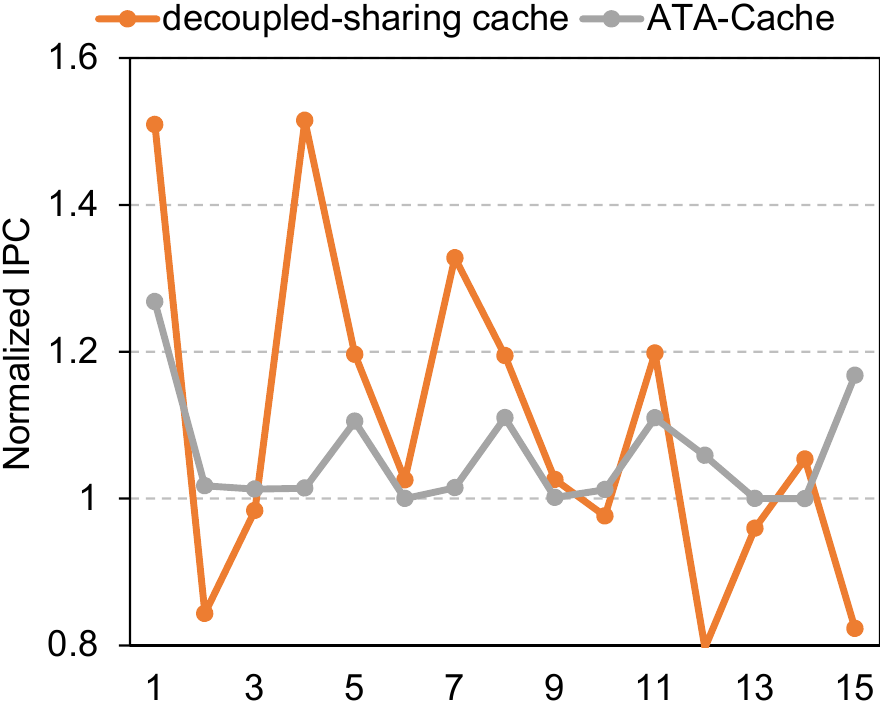}}
	\subfigure[conv3d]{
		\label{kernel performance conv3d}
		\includegraphics[width=0.48\linewidth]{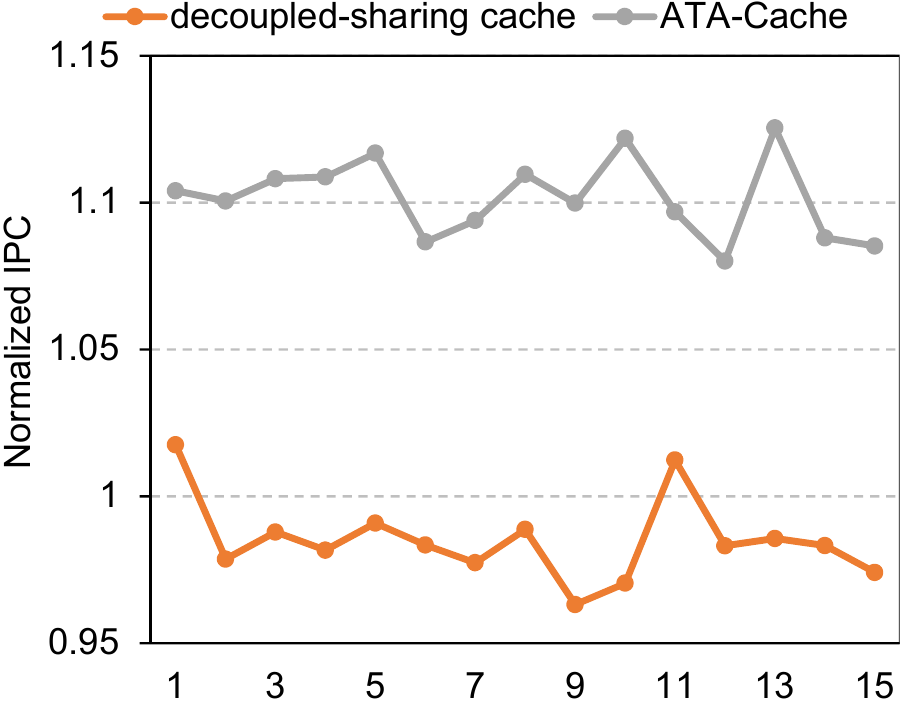}}
	\subfigure[HS3D]{
		\label{kernel performance hotspot3d}
		\includegraphics[width=0.48\linewidth,height=3.4cm]{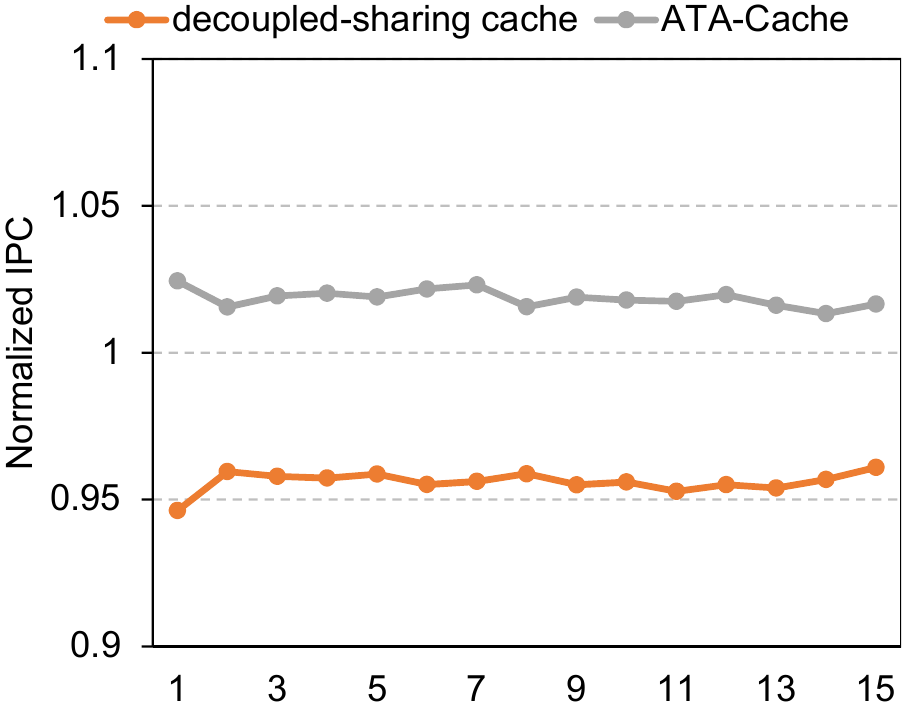}}
    \subfigure[sradv1]{
		\label{kernel performance sradv1}
		\includegraphics[width=0.48\linewidth,height=3.4cm]{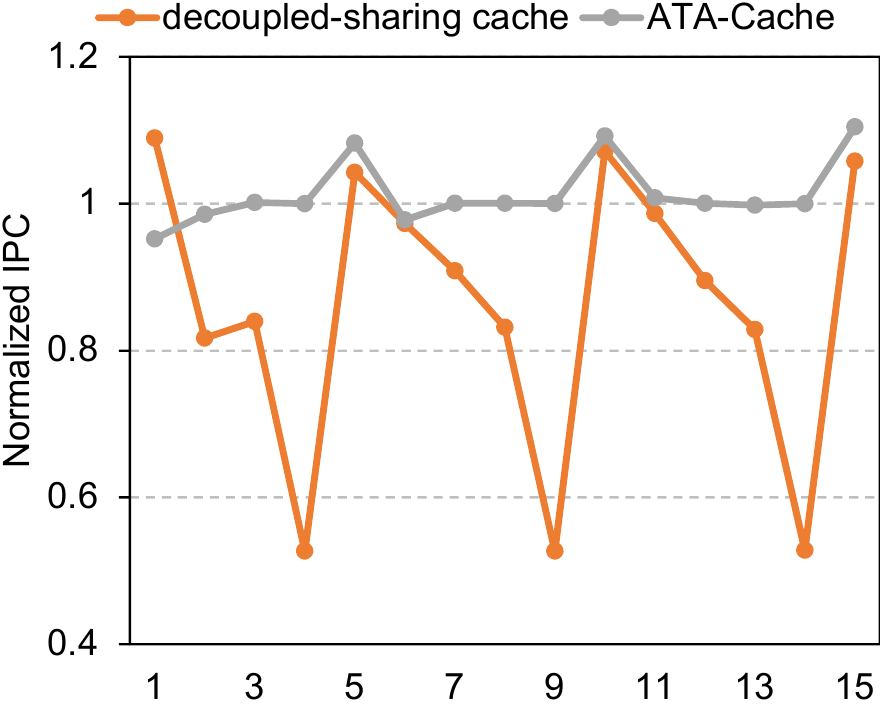}}
	\caption{Performance of kernels in four GPU applications (normalized to private cache).}
	\label{kernel performance}
 \vspace{-1em}
\end{figure}

A GPU application usually consists of multiple kernels. These kernels have access diversity, resulting in various performance improvements for different kernels in the same application. As shown in Figure \ref{kernel performance}, we choose two high inter-kernel locality applications(SN and conv3d) and two low inter-kernel locality applications(HS3D and sradv1) to study the performance of each kernel in the application. As shown in Figure \ref{kernel performance SN}, for SN, the ATA-Cache performance improvement is lower than that of the decoupled-sharing cache on some kernels. However, the decoupled-sharing cache degrades performance on multiple kernels due to high contentions, so the overall performance of ATA-Cache is better than that of the decoupled-sharing cache. Figure \ref{kernel performance conv3d} and Figure \ref{kernel performance hotspot3d} show that the ATA-Cache outperforms the decoupled-sharing cache on all kernels for both HS3D and conv3d. In Figure \ref{kernel performance sradv1}, for kernel 4, kernel 9, and kernel 14, the decoupled-sharing cache performance is significantly degraded, resulting in a reduction in overall performance.
In summary, ATA-Cache can effectively improve IPC for applications with a high inter-core locality. For applications with poor inter-core locality, IPC is significantly better than decoupled-sharing cache due to fewer contentions.

\subsection{L1 Cache Latency}\label{4C}
\begin{figure}[t]
\centerline{\includegraphics[width=0.5 \textwidth]{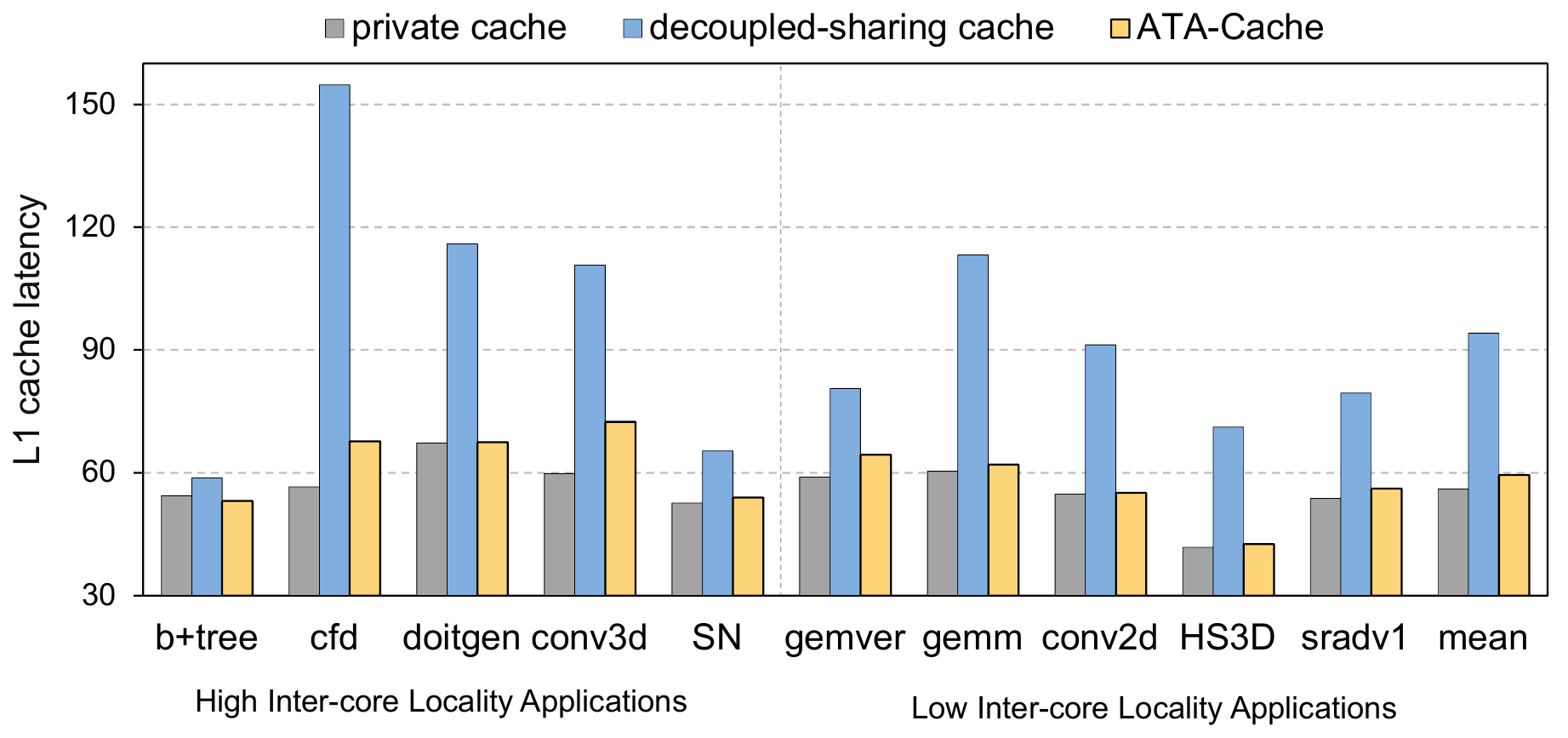}}
\caption{Illustrating the benefits of the ATA-Cache in terms of L1 cache latency.}
\label{Fig: ins}
\vspace{-1em}
\end{figure}

We also evaluate the L1 access latency, which is the completion time of the L1 cache accesses for all requests from a single load instruction. 
Figure \ref{Fig: ins} shows the experimental results of the L1 cache latency. 
Higher L1 cache latency means more severe resource contentions due to sharing, which leads to GPU performance degradation. The decoupled-sharing cache increases the L1 cache latency by 67.2\% on average (up to 2.74x) over the private cache in both high inter-core locality applications and low inter-core locality applications. ATA-Cache reduces resource contentions due to sharing, so L1 cache latency increases by only 6.0\% over the private cache. To sum up, ATA-Cache makes full use of inter-core locality while introducing much fewer additional resource contentions than decoupled-sharing cache, resulting in more performance gains.

\subsection{Hardware Overhead}

The main extra area and power overhead of the special design in ATA-Cache come from the crossbar and comparator groups used in the aggregated tag array. We evaluate the hardware overhead assuming a 45nm technology\cite{Nanjia}. Crossbar and comparator groups lead to area overhead of 1.02 $mm^{2}$ and 0.02 $mm^{2}$ and the total leakage power overhead is 5.55 $mw$. 
\section{Conclusion}
In this article, we discuss the serious resource contention problem in the GPU shared L1 cache. We propose ATA-Cache, a shared L1 cache design with an aggregated tag array that dramatically reduces resource contentions compared to the conventional GPU shared cache. We decouple the tag arrays from the GPU L1 caches and aggregate them into an aggregate tag array. In the aggregated tag array, the request address tags can be compared in parallel with the tag arrays of all L1 caches, thus probing and leveraging the replicated data with few contentions. The L1 cache latency of ATA-Cache is significantly lower than that of the decoupled-sharing cache, and the GPU performance of ATA-Cache is greatly improved.

\bibliographystyle{IEEEtr}
\bibliography{mybib}

\begin{thebibliography}{10}
\providecommand{\url}[1]{#1}
\csname url@samestyle\endcsname
\providecommand{\newblock}{\relax}
\providecommand{\bibinfo}[2]{#2}
\providecommand{\BIBentrySTDinterwordspacing}{\spaceskip=0pt\relax}
\providecommand{\BIBentryALTinterwordstretchfactor}{4}
\providecommand{\BIBentryALTinterwordspacing}{\spaceskip=\fontdimen2\font plus
\BIBentryALTinterwordstretchfactor\fontdimen3\font minus
  \fontdimen4\font\relax}
\providecommand{\BIBforeignlanguage}[2]{{%
\expandafter\ifx\csname l@#1\endcsname\relax
\typeout{** WARNING: IEEEtran.bst: No hyphenation pattern has been}%
\typeout{** loaded for the language `#1'. Using the pattern for}%
\typeout{** the default language instead.}%
\else
\language=\csname l@#1\endcsname
\fi
#2}}
\providecommand{\BIBdecl}{\relax}
\BIBdecl

\bibitem{choukse2020buddy}
E.~Choukse \emph{et~al.}, ``Buddy compression: Enabling larger memory for deep
  learning and hpc workloads on gpus,'' in \emph{Proc. of ISCA}, 2020, pp.
  926--939.

\bibitem{kwon2021tensor}
Y.~Kwon \emph{et~al.}, ``Tensor casting: Co-designing algorithm-architecture
  for personalized recommendation training,'' in \emph{Proc. of HPCA}, 2021,
  pp. 235--248.

\bibitem{sun2022gtuner}
Q.~Sun \emph{et~al.}, ``Gtuner: tuning dnn computations on gpu via graph
  attention network,'' in \emph{Proc. of DAC}, 2022, pp. 1045--1050.

\bibitem{kim2018gpu}
M.~Kim \emph{et~al.}, ``A gpu-aware parallel index for processing
  high-dimensional big data,'' \emph{IEEE Transactions on Computers}, vol.~67,
  no.~10, pp. 1388--1402, 2018.

\bibitem{wang2022a2}
Q.~Wang \emph{et~al.}, ``A2-ilt: Gpu accelerated ilt with spatial attention
  mechanism,'' in \emph{Proc. of DAC}, 2022, pp. 967--972.

\bibitem{wulf1995hitting}
W.~A. Wulf \emph{et~al.}, ``Hitting the memory wall: Implications of the
  obvious,'' \emph{ACM SIGARCH computer architecture news}, vol.~23, no.~1, pp.
  20--24, 1995.

\bibitem{li2015inter}
D.~Li \emph{et~al.}, ``Inter-core locality aware memory scheduling,''
  \emph{IEEE Computer Architecture Letters}, vol.~15, no.~1, pp. 25--28, 2015.

\bibitem{dublish2016cooperative}
S.~Dublish \emph{et~al.}, ``Cooperative caching for gpus,'' \emph{ACM
  Transactions on Architecture and Code Optimization (TACO)}, vol.~13, no.~4,
  pp. 1--25, 2016.

\bibitem{ibrahim2019analyzing}
M.~A. Ibrahim \emph{et~al.}, ``Analyzing and leveraging remote-core bandwidth
  for enhanced performance in gpus,'' in \emph{Proc. of PACT}, 2019, pp.
  258--271.

\bibitem{ibrahim2020analyzing}
M.~A. Ibrahim \emph{et~al.}, ``Analyzing and leveraging shared l1 caches in
  gpus,'' in \emph{Proc. of PACT}, 2020, pp. 161--173.

\bibitem{ibrahim2021analyzing}
M.~A. Ibrahim \emph{et~al.}, ``Analyzing and leveraging decoupled l1 caches in
  gpus,'' in \emph{Proc. of HPCA}, 2021, pp. 467--478.

\bibitem{zhao2019adaptive}
X.~Zhao \emph{et~al.}, ``Adaptive memory-side last-level gpu caching,'' in
  \emph{Proc. of ISCA}, 2019, pp. 411--423.

\bibitem{tripathy2021paver}
D.~Tripathy \emph{et~al.}, ``Paver: Locality graph-based thread block
  scheduling for gpus,'' \emph{ACM Transactions on Architecture and Code
  Optimization (TACO)}, vol.~18, no.~3, pp. 1--26, 2021.

\bibitem{baruah2020valkyrie}
T.~Baruah \emph{et~al.}, ``Valkyrie: Leveraging inter-tlb locality to enhance
  gpu performance,'' in \emph{Proc. of PACT}, 2020, pp. 455--466.

\bibitem{li2019efficient}
B.~Li \emph{et~al.}, ``An efficient gpu cache architecture for applications
  with irregular memory access patterns,'' \emph{ACM Transactions on
  Architecture and Code Optimization (TACO)}, vol.~16, no.~3, pp. 1--24, 2019.

\bibitem{khairy2020accel}
M.~Khairy \emph{et~al.}, ``Accel-sim: An extensible simulation framework for
  validated gpu modeling,'' in \emph{Proc. of ISCA}, 2020, pp. 473--486.

\bibitem{che2009rodinia}
S.~Che \emph{et~al.}, ``Rodinia: A benchmark suite for heterogeneous
  computing,'' in \emph{Proc. of IISWC}, 2009, pp. 44--54.

\bibitem{karki2019tango}
A.~Karki \emph{et~al.}, ``Tango: A deep neural network benchmark suite for
  various accelerators,'' in \emph{Proc. of ISPASS}, 2019, pp. 137--138.

\bibitem{Polybench}
L.-N. Pouchet and S.~Grauer-Gray, ``Polybench: The polyhedral benchmark
  suite.'' \url{http://web.cs.ucla.edu/\~pouchet/software/polybench/}, 2012.

\bibitem{Nanjia}
``Nangate inc. nangate 45nm open cell library,'' \url{http://www.nangate.com},
  2008.

\end{thebibliography}

\vspace{12pt}
\color{red}

\end{document}